\documentclass[authorversion,acmlarge]{acmart}

\acmYear{2020}\copyrightyear{2020}
\setcopyright{acmcopyright}
\acmConference[CryBlock 2020]{3rd Workshop on Cryptocurrencies and Blockchains for Distributed Systems}{September 25, 2020}{London, United Kingdom}
\acmBooktitle{3rd Workshop on Cryptocurrencies and Blockchains for Distributed Systems (CryBlock 2020), September 25, 2020, London, United Kingdom}
\acmPrice{15.00}
\acmDOI{10.1145/3410699.3411361}
\acmISBN{978-1-4503-8079-9/20/09}

\begin{document}

\title{DiLeNA: Distributed Ledger Network Analyzer}

\author{Luca Serena}
\email{luca.serena2@unibo.it}
\affiliation{%
  \institution{CIRI ICT}
  \city{Cesena}
  \country{Italy}
}

\author{Stefano Ferretti}
\email{stefano.ferretti@uniurb.it}
\affiliation{%
  \institution{University of Urbino}
  \city{Urbino}
  \country{Italy}
}

\author{Gabriele D'Angelo}
\email{g.dangelo@unibo.it}
\affiliation{%
  \institution{University of Bologna}
  \city{Cesena}
  \country{Italy}
}


\begin{abstract}
This paper describes the Distributed Ledger Network Analyzer (DiLeNA), a new software tool for the analysis of the transactions network recorded in Distributed Ledger Technologies (DLTs). The set of transactions in a DLT forms a complex network. Studying its characteristics and peculiarities is of paramount importance, in order to understand how users interact in the distributed ledger system. The tool design and implementation is introduced and some results are provided. In particular, the Bitcoin and Ethereum blockchains, i.e.~the most famous and used DLTs at the time of writing, have been analyzed and compared.
\end{abstract}




\maketitle

\section{Introduction}
The blockchain is a relatively young technology and the number of people using cryptocurrencies as well as other services and applications, that are based on such a data structure, is constantly growing \cite{users,ieeeaccess2020}. The various systems are based on different design choices and there might be several motivations to be interested in cryptomarkets. For example, some users might be interested in the anonymity (or pseudo-anonymity) features provided by these technologies \cite{michael2018blockchain}, others in the lack of central entities in charge of managing the money transfers, in the economic value of the cryptocurrencies built over these ledgers (for example, for speculation purposes), or finally, with the aim to build decentralized applications that are able to exploit the features offered by smart contracts \cite{luu2016making}.

Blockchains and cryptocurrencies have been widely studied in terms of security issues \cite{smart-sec}. Similarly, their economical impact received attention \cite{li2017technology}. However, it could be also interesting to study these technologies using mechanisms that have been applied in the social networks analysis (e.g. Facebook \cite{ugander2011anatomy}, Twitter \cite{gabielkov2014studying}). For example, studying the interactions among the different accounts involved in the transactions that are recorded on the Distributed Ledger Technologies (DLTs). For this purpose, the system can be seen as a graph where links are drawn when pairs of nodes interact. Unlike normal methods for exchanging money, here the transactions are visible to every user, so this kind of analysis is viable \cite{sf-gda}.

In this paper, we present a new tool called DiLeNa (Distributed Ledger Network Analyzer) that has been developed to analyze the interactions among either Bitcoin or Ethereum accounts, by using real data stored on the respective DLTs. Specifically, the goal of the software is to provide some information on the structuring of the analyzed networks and to observe if the graph representing the interactions among the accounts has small world properties.

Our analysis shows that, in both Bitcoin and Ethereum networks, there is a vast percentage of nodes whose total degree is equal to $1$, meaning that they were involved in some interaction (corresponding to the exchange of cryptos, or smart contracts invocation) with a single other node. Both networks have very few hubs in their networks, i.e.~very active nodes that make many transactions with a large set of nodes. In fact, for both the networks there is a negligible number of nodes (around $10$) that interacted, at most, with approximately the $2\%$ of the network node set. In the case of Ethereum, a single node had connections with approximately the $10\%$ of the network set. In view of the fact that the considered time period for the analysis of the blockchain is relatively short, i.e.~one month, this is a non-negligible aspect. As concerns the small world property, our analysis shows that (in the considered period) the two networks have different characteristics, i.e.~Ethereum is a small world, while Bitcoin is not.

The remainder of the paper is organized as follows. Section 2 introduces some background and related work. Section 3 describes the design choices of the program and deals with the critical aspects of the implementation that need to be taken in account. Section 4 details the structure of the software. Section 5 analyzes the results obtained analyzing the Ethereum and Bitcoin blockchains. Finally, Sections 6 provides some concluding remarks.

\section{Background and Related Work}\label{sec:back}
In this section, some background that is essential to understand the rest of the paper is introduced. Specifically, the topics covered will be the DLT technology, the representation of complex systems as graphs and some metrics that can be used in order to evaluate if a network has small world properties.

\subsection{DLT and Blockchain Technologies}
The DLT is a data structure that works as a distributed and immutable database. Several nodes keep track of the entire data and check the integrity of the records, that is guaranteed by cryptographic functions. A DLT is defined as permissionless if it is not needed any prior approval to actively participate to the system.

Blockchains are a specific kind of DLTs, where the decentralized ledger is organized as a chain of blocks containing data. Such data are often referred as ``transactions'', since these ledgers have been originally devised to track cryptocurrency exchange (e.g.~Bitcoin). In permissionless blockchains like Bitcoin, each user can read and add data, while there is no way to cancel information. Moreover, there is always a protocol (called consensus algorithm) that allows all the nodes to agree about which data have to be considered valid.

The DLT is the underlying technology of most the cryptocurrencies. The first as well as the most famous cryptocurrency is Bitcoin, that was created in 2008 \cite{nakamoto2019bitcoin}. Recently, also Ethereum gained popularity because it allows to execute, other than simple transactions, actual contracts written with code, the so called ``smart contracts''. In both Ethereum and Bitcoin, the users are identified with a public cryptographic key and there is no trivial way to associate the public keys with the real identity of the users. So it is possible (and often happens) that certain users control multiple accounts.

\subsection{Graphs and Complex Systems}
A graph $G$ is a data structure defined as $G=(V,E)$ where $V$ is a set of nodes (vertices) and $E$ is a set of edges that link a pair of nodes. Graphs can have different configurations: they can either be directed or undirected and they can either be weighted or not. In an undirected graph, an edge links two nodes in both directions, that means that if $A$ is connected with $B$ then $B$ is also connected with $A$. This condition is not necessarily true in directed graphs. A weight is a numerical value associated with an edge: it can represent, depending on the meaning of the graph, a distance, a cost and so on. In other cases, like for social graphs, there is no meaning for a weight, and all we want to know is whether two vertices are connected or not. In a graph, it is not always possible to reach each couple of nodes following a path along the edges. A subset of a graph in which any couple of nodes is connected by a path is called ``connected component''. The component with the greatest number of vertices is often referred as the ``main component''.

Depending on the problem, the meaning of a graph can be different. In a map, for example, vertices may represent some locations and the weights of the edges are the distance between two nodes \cite{sf-gda,ferretti2017modeling,FERRETTI20131631}. In our case, the vertices are the active users of a certain blockchain and the presence of an edge indicates that there has been an interaction between the two nodes.

\subsection{Small World and Random Graphs}
The random graphs are networks where there is a predefined number of nodes and edges and all the links are assigned randomly. A strategy to generate a random graph involves the usage of the Erdos-Renyi model \cite{paul1959random}: a graph is created by selecting as parameters the total number of nodes and either the total number of edges or the probability that two nodes are connected between them.

Small world networks \cite{watts1998collective} are a graph topology where most of the nodes are not connected to each other, but however, most of the vertices can be reached by other nodes through a short path. In these graphs, the vertices tend to form clusters and the average distance between two random nodes is usually minor, with respect to other graphs topologies, like random graphs. Detecting if a graph has this property can be useful in different areas. For example, in medicine it can give information about how a disease spreads among the people, in computer science it would be possible to exploit this knowledge in order to optimize the dissemination and the storage of data around the network \cite{gda-jpdc-2017}.

There are some metrics used to evaluate if a graph has small world properties \cite{bassett2006small}: 
\begin{itemize}
    \item \emph{Average shortest path length}, that is, the average among all the shortest paths between two nodes of the graph belonging to the same component. It can be calculated using the Dijkstra's algorithm \cite{dijkstra1959note}, in this case the time complexity of the algorithm is $O(|V||E|+|V|^2 log|V|)$.
    \item \emph{Average clustering coefficient}, that is, the average of the clustering coefficients of all the nodes. It produces an output ranging from $0$ to $1$, the higher is the number the more clustered is the graph. The clustering coefficient of a node is the fraction that indicates how many edges between his neighbors exist among all the possible ones.
\end{itemize}
After getting the outcome of the metrics, a comparison with a random graph with the same number of nodes and edges is needed. One can state that the graph features small world properties if, compared to a random graph of equal size, the average clustering coefficient is significantly higher and the average shortest path length similar (or smaller) \cite{ferretti2017modeling}.

\section{The DiLeNa Tool}
The proposed software, that is freely available on the research group website \cite{pads}, is modular and it is structured as two phases:
\begin{itemize}
    \item \emph{Graph Generator}: the transactions contained in a certain number of blocks (of either Ethereum or Bitcoin) are downloaded and an undirected graph representing the interactions among the nodes of the system is created. The vertices of the graph correspond to the accounts in the DLT and for each transaction an edge between the two involved users is made (if not already existing).
    \item \emph{Graph Analyzer}: some typical metrics of the obtained graph are calculated. For instance, the tool measures the degree distribution, the network clustering coefficient, the identification of the main component and some of its main metrics, such as the average shortest path. Then, the analyzed graph is compared with a random generated graph with the same amount of nodes and edges, in order to understand if the network is a small world.
\end{itemize}

The generated graph is stored in two different formats: JSON and Pajek. More specifically, in JSON the vertices are encoded as a list of hexadecimal strings representing the public keys of the users and the edges are triples composed of sender's key, recipient's key and the amount of cryptocurrency transferred. In Pajek, the encoding of vertices is based on ID-keys and in the representation of the edges senders and receivers are indicated using the vertices' IDs instead of the key. This approach permits to save a lot of space in memory since the representation of the nodes using integer IDs requires fewer bytes than the respective public keys (that are strings of about 40 characters each).

\subsection{Graph Generator}
This module of DiLeNa is implemented using the JavaScript programming language in conjunction with Node.js, a run-time environment that allows to execute JavaScript code outside a web browser.

The main technologies used in the implementation of the Graph Generator are:
\begin{itemize}
    \item \emph{Web3.js}: a collection of libraries used for interacting with a local or remote Ethereum node. In our case, it is used to download Ethereum transactions. Since an analogous service for Bitcoin was not available (at least, to the best of our knowledge at the time of the software implementation), the \textit{libcurl} file transfer library was used to download transactions from the site \textit{blockchain.info};
    \item \emph{Infura}: a service that provides access to a remote Ethereum node though Application Programming Interfaces (APIs). Users can generate application keys with whom they can connect to the blockchain without having to download the entire distributed ledger and managing its synchronization. Unfortunately, recent restrictions on the number of actions that can be made slowed down the download process and forced the implementation in DiLeNa of a recovery system. In fact, if a request for a block is rejected or if it results in an error, then such a request is automatically repeated.
\end{itemize}

The Graph Generator module is structured in two parts:
\begin{itemize}
    \item \emph{Live}: this part is used to download the transactions contained in blocks that are generated while DiLeNa is running. In other words, it is aimed to collect the new transactions in real-time. In this case, both a complete transactions graph and partial graphs are created. The later are used by a web-based graph rendering module (called Ethereum Galaxy) that is able to live stream the graph evolution both in terms of nodes and links.
    \item \emph{Offline}: this part is thought to deal with large amounts of data. In fact, the user can select a time interval and all the corresponding blocks are downloaded from the DLT. In practice, this permits the analysis of transactions that have been produced sometimes in the DLT history. This approach can be very useful for studying the evolution of DLTs during their lifespan. For example, to compare the topology structure of Bitcoin in the first years after its development with the current network structure (at the peak of its popularity).
    \end{itemize}

\subsection{Graph Analyzer}
This part of DiLeNa has been specifically implemented in Python to interoperate with the NetworkX library, a software library for the analysis and the manipulation of complex networks\cite{networkx}.

Under the usage viewpoint, the Graph Analyzer enables the user specifying what graph is to be analyzed and then to compute some specific metrics:
\begin{itemize}
    \item Degree Distribution;
    \item Average Clustering Coefficient (ACC);
    \item Average Shortest Path Length (ASPL) of the main component.
\end{itemize}
After completing these computations, the Graph Analyzer generates a new random graph that has the same number of nodes (and an equivalent number of edges) than the graph that has been just analyzed. This allows to make a comparison between the two graphs, in order to note if the studied network has small world properties.

\subsection{Critical Aspects}
The most critical aspect of downloading a large number of DLT blocks, reconstructing a graph data structure and then computing some specific metrics on the graph, concerns with the efficiency of the full procedure (e.g.~time minimization) and the reduction of the amount of memory used in the whole process (i.e~memory management).

With the aim to boost the efficiency, a good strategy is to use a parallel approach, in all parts of the software, when it is possible. The graph creation process employs two levels of parallelization, one for downloading the transactions and another for building the interactions graph. The logic for the parallelization of the download follows a coordinator-workers approach. The coordinator assigns some tasks to the workers and each worker downloads the transactions of the blocks with the given indices and then saves them in a temporary file. When a worker completes the assigned tasks, then it contacts again the coordinator in order to get new tasks. In most cases, the maximum efficiency of this approach is obtained when the number of workers is set as the number of cores of the CPU, but both the network data transfer bottleneck and the bandwidth of the data storage must be considered.

For which concerns the parallelization of the graphs generation, two different threads are employed to create the JSON and the Pajek files. More specifically, each graph generation thread concurrently reads the data from the temporary files generated during the download phase. In this specific case, there are no concurrency issues since the access to the shared resources is read-only.

Finally, DiLeNa implements some level of parallelization, when possible, also in the analysis phase. Again a coordinator-workers approach is implemented: the coordinator is in charge of assigning to the workers the specific graph nodes on which to calculate the metrics and the number of tasks corresponds to the total number of nodes in the network.

Another critical aspect that must be considered in DiLeNa is the memory management. In fact, large graphs are not suitable for being saved as a whole in the host random-access memory. For this reason, it is necessary to temporarily store several data structures (representing different parts of the graph) in secondary memory. Read and write operations from/to the mass storage are more expensive than main memory accesses; however, the advantage of this approach is that it permits to deal with large graphs, without being constrained by the hardware limitations (e.g.~the amount of free main memory available in the host used for running the software tool).

The download process of the blocks from the DLT can last several hours or even days, depending on how long is the time-span that the user has decided to analyze. For example, a user could be willing to download all blocks generated during a specific month, an year or even a larger time interval (e.g. the whole DLT history). Thus, it is important to be able to pause the download and to resume it at a later time. Due to the already mentioned Infura restrictions in the APIs usage, it is also necessary to manage the case where the download of a block is not successful: in that case, the request is repeated until it is satisfied. Because of this constraint, in the usage of DiLeNa, the download phase can last up to 100 times more than the construction of the graph. The analysis of the graph is in turn costlier in terms of time and it can require up to several days or some weeks to be completed (on an average desktop computer), depending on the size and on the structuring of the graph. The amount of time that is necessary to analyze the generated graph can be greatly reduced using host equipped with a large number of CPU cores or employing High Performance Computing (HPC) architectures.

An alternative approach for downloading the data to be analyzed is to mirror the entire DLTs. This permits to avoid the usage of Infura APIs (or libcurl) and therefore to overcome the imposed limitations but it has also relevant drawbacks. In fact, the blockchain size for Ethereum is now in the order of 4 terabytes and of 270 gigabytes for Bitcoin\cite{size}. It is clear that this approach should be pursued only in specific cases, for example when it is expected that large parts of the blockchain will have to be analyzed.

\section{Source Code Structuring}
This section describes how the DiLeNa source code is structured in packages.

\subsection{Graph Generator}
This part of the DiLeNa is composed of seven packages:
\begin{itemize}
    \item \emph{live}: it manages the live download. After it has been established a connection with the data provided (e.g. the Infura APIs), it periodically checks if new blocks have been added to the DLT and, in case, the transactions contained in the new blocks are downloaded. Furthermore, some additional files that are used for the graphical visualization of the interaction network are created;
    \item \emph{no-layout}: it contains all the modules that are necessary for the offline mode. After input validation, the time interval is used to identify the corresponding data blocks (or, depending on the specific DLT, data pointer). Then, the download can finally start. Furthermore, if the user chooses to download the whole blockchain then it is checked if a previous copy has been previously downloaded, in order to identify and to request only the missing data.
    \item \emph{download}: it contains all the logic to download the transactions according to the coordinator-workers approach;
    \item \emph{generation}: it manages the creation of the graphs, and embodies the functionalities to export it in some specific file formats. The graphs are generated from the downloaded transactions that are locally stored;
    \item \emph{shutdown}: it is activated when the graph generation process or the download phase must be interrupted, for some reason. The current state is saved so that is possible to resume the processes at a later time.
    \item \emph{utilities}: it embodies the main support functions, used in different parts of the software;
    \item \emph{current-project-state}: it is used to configure the different DLTs, and the graph formats, that are currently supported in DiLeNa.
\end{itemize}

\subsection{Graph Analyzer}
This module is smaller, if compared to the other part of the software. In fact, in this case all functionalities are implemented within a single package. The \textit{Main} module is in charge of managing and coordinating all the task that need to be completed. This module is used to select the metrics that need to be calculated on the graph, it triggers the loading of the graph, the parallel computation of the metrics on the generated graph and the creation of the corresponding random graph (as described above). After the computations are finished, it compares the results obtained from the two graphs and then and it stores the resulting data on persistent storage.

\section{Analysis of the Results}
In this section, the outcomes of our analyses are discussed. In particular, we have analyzed a full month of interactions of both blockchains, specifically December 2010 for Bitcoin, and December 2016 for Ethereum. These specific time frames have been selected with the aim to compare the Ethereum and Bitcoin blockchains at the same stage of evolution, instead of comparing a mature technology with a newly introduced one. For this reason, for both of them, it has been decided to take into account the last month of the second year the blockchain was active.

Since in six years the use of DLT-based technologies has considerably increased, it turned out that Ethereum graph is more than five times bigger than Bitcoin graph. Cryptocurrencies were still not a well-known tool in 2010, this probably explains the size difference. It could have been interesting to consider a longer period, but the time required to calculate the metrics, especially the average shortest path length, was a strong limitation. For example, in the hardware used for the experiments, the computations to get the metrics results on the Ethereum graph took more than two weeks to be completed.

To state that a graph has small world properties, it is required that the ratio of the average clustering coefficients between the analyzed graph and the corresponding random graph must be significantly greater than $1$. On the other hand, for which concerns the average shortest path length, the ratio should be equal or less than $1$.

In the following, we report the results that we obtained when analyzing both the Ethereum and the Bitcoin blockchains. 

\begin{table*}[h]
  \centering
  \caption{Main metrics of the considered blockchain networks.}
\begin{tabular}{|c|c|c|c|c|}
\toprule
\textbf{Blockchain}   & \textbf{net size} & \textbf{net avg degree} & \textbf{main component size} & \textbf{main component avg degree} \\
\midrule
\textit{Ethereum}      & 111633 & 3.5 & 96.8\% net size & 3.6 \\
\textit{Bitcoin}     & 21339  & 2.7 & 86.8\% net size & 2.7  \\
\bottomrule
\end{tabular}
\label{Table:confronto}
\end{table*}

Table \ref{Table:confronto} shows a first comparison of the main metrics of the considered blockchain networks. As already mentioned, the Bitcoin network is smaller than the Ethereum one. In fact, while the Ethereum net was in the order of $100K$ nodes, the Bitcoin network size is about $20K$ nodes. This is simply explained by the fact that Ethereum in 2016 had more active users than Bitcoin in 2010. The amount of nodes in the networks corresponds to addresses. In the case of Bitcoin, these addresses correspond to specific users (i.e. a user can have multiple addresses), while in Ethereum addresses can be associated to users (external addresses) as well as to smart contracts (internal addresses). Thus, the network counts also for active smart contracts that were triggered through external transactions from users.

\subsection{Ethereum}
The degree distribution resulting from our analysis shows that there are few nodes with a very high degree, thus implying the presence of hubs. In particular, a negligible number of nodes (around $10$) has a degree higher than $2000$, meaning that they had some interactions with the $2\%$ of the node set. The node with the highest degree showed an amount of connections with almost the $10\%$ of the node set.

\begin{figure}[t]
  \centering
    \includegraphics[width=0.7\textwidth]{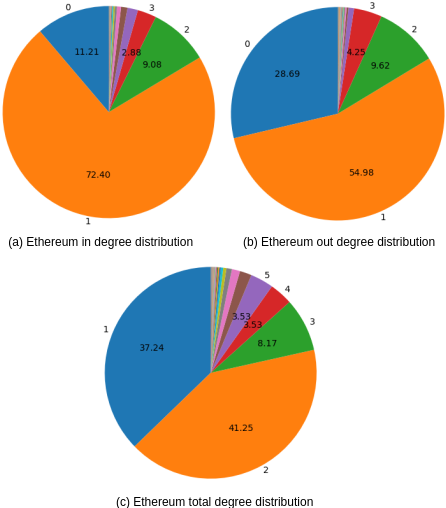}
  \caption{In, out and total degree distribution of the Ethereum graph.}
  \label{ethdegree}
\end{figure}

It is also possible to identify a big percentage of active nodes who either did not send or receive any transaction (respectively the $28.7\%$ and the $11.2\%$ of all nodes in the network). In other words, most nodes participated as senders or recipients of a single transaction, in the considered time period. Thus, their in-degree is greater than or equal to $1$ and out-degree is $0$, or vice versa. Figure \ref{ethdegree} shows some pie charts representing the degree distribution of the Ethereum network (respectively in-degree, out-degree and total degree).

\begin{table*}[h]
  \centering
  \caption{Comparison between the Ethereum graph and the equivalent random graph.}
\begin{tabular}{|c|c|c|c|}
\toprule
\textbf{Graph}   & \textbf{Graph ACC} & \textbf{Main Component ASPL} & \textbf{Main Component ACC} \\
\midrule
\textit{Ethereum}      & 0.02099 & 1.4256 & 0.02134 \\
\textit{Random}      & 0.000014 & 10.3584 & 0.000015  \\
\bottomrule
\end{tabular}
\label{ethtable}
\end{table*}

The outcome of the other metrics that have been computed on the Ethereum network is shown in Table \ref{ethtable}. For what regards the comparison with the corresponding random graph, the ratio of the average clustering coefficients is $1469$ and the ratio of the average shortest path lengths is $0.14$. This permits us to state that the Ethereum network has small world properties.

\subsection{Bitcoin}
\begin{figure}[t]
  \centering
    \includegraphics[width=0.7\textwidth]{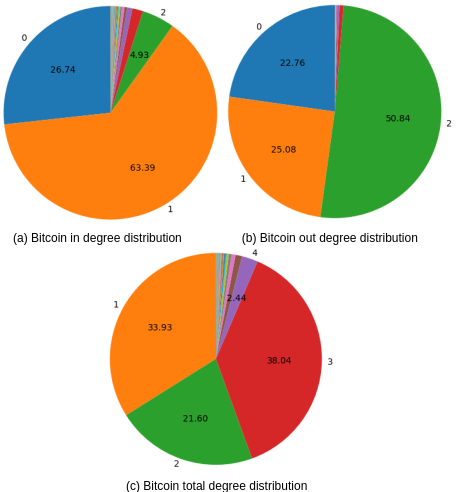}
  \caption{In, out and total degree distribution of the Bitcoin graph.}
  \label{bitdegree}
\end{figure}

Also in this case, the analysis of the Bitcoin degree distribution shows the presence of very few hubs. Similarly to Ethereum, these hubs were connected to approximately the $2\%$ of the node set. In this case, $22.76\%$ of nodes did not make any transaction (out-degree equal to $0$) and the $26.74\%$ never received any amount of cryptocurrency (in-degree equal to $0$). Less than $4\%$ of the vertices participated in $5$ or more transactions. Only few nodes are characterized by a very high degree, of the order of hundreds of connections (i.e. hub nodes). Figure \ref{bitdegree} shows some pie charts representing the degree distribution of the Bitcoin network.

\begin{table*}[h]
  \centering
  \caption{Comparison between the Bitcoin graph and the equivalent random graph.}
\begin{tabular}{|c|c|c|c|}
\toprule
\textbf{Graph}   & \textbf{Graph ACC} & \textbf{Main Component ASPL} & \textbf{Main Component ACC} \\
\midrule
\textit{Bitcoin}    & 0.0235 & 190.4879 & 0.024 \\ 
\textit{Random}    & 0.000026 & 6.461 & 0.000029  \\
\bottomrule
\end{tabular}
\label{btctable}
\end{table*}

Table \ref{btctable} shows the outcome of the other metrics that have been computed on the Bitcoin network. Also in this case, the ratio of the average clustering coefficients is high (i.e. $828.3$), although it is lower if compared to what obtained for Ethereum. However, in the case of Bitcoin, the network does not show small world properties since the ratio of the average shortest path lengths is $29.5$.

\section{Conclusions}\label{sec:conc}
In this paper, we present the Distributed Ledger Network Analyzer (DiLeNA), a new software tool designed for downloading the transactions recorded in the Distributed Ledger Technologies (DLTs) and to compute some metrics on the resulting interaction graph. The current version of DiLeNa supports Bitcoin and Ethereum, enabling the analysis of other DLTs is straightforward but is left as a future work.

The results reported in the paper, show that the Ethereum network has small world properties but the same does not occur in Bitcoin. In our opinion, this happens due to the presence of smart contracts in Ethereum (that are not available in Bitcoin). In fact, many interactions among groups of users are performed through smart contracts, that thus become common network neighbors to all these users. Furthermore, in both blockchains users can control multiple accounts. Indeed, this is a common practice adopted by many Bitcoin wallets, i.e.~for any novel transaction, a new address is generated, in order to increase the users anonymity and unlinkability between transactions (or at least, reduce the ease to aggregate accounts and de-anonymize them). This surely lowers the nodes' average degree, the network clustering coefficient and increases the average number of hops needed to connect two nodes.

As reported before, the modular structure of DiLeNa permits to easily add the support for other DLTs, such as IOTA, Ripple and Litecoin. 
Another future extension of DiLeNa aims to further increase the computation parallelization thought GPUs. In this case, the main bottleneck is the lack of support for GPUs in the current version of the NetworkX library used for computing some of the metrics described in the paper. In other words, this will require us to switch to another library or to embed the metrics computation in DiLeNa using a more efficient programming language.

\section*{Acknowledgements}
The authors of this paper would thank Dr. Pietro Di Lena for unwittingly providing the best acronym for the DLT network analyzer (i.e. DiLeNa) and Daniele Rosetti for working on the implementation of the software tool.

\bibliographystyle{ACM-Reference-Format}
\bibliography{biblio}

\end{document}